\documentclass[sigconf]{acmart}
\renewcommand\footnotetextcopyrightpermission[1]{}
\usepackage[boxruled,noend]{algorithm2e}
\usepackage{multirow}
\usepackage{subcaption}

\graphicspath{{fig/}}
\newcommand{\Ali}{\texttt{Alibaba}\xspace}
\newcommand{\VE}{\texttt{Visual Encoder}\xspace}
\newcommand{\CTR}{\texttt{CTR predictor}\xspace}
\newcommand{\vfi}{\mathbf{v}^{S1}}
\newcommand{\vft}{\mathbf{v}^{S2}}
\newcommand{\vfd}{\mathbf{v}^{D}}
\newcommand{\vf}{\mathbf{v}}
\newcommand{\Real}{\mathbb{R}}
\newcommand{\Loss}{\mathbb{L}}
\acmConference[Woodstock '18]{Woodstock '18: ACM Symposium on Neural
  Gaze Detection}{June 03--05, 2018}{Woodstock, NY}
\acmBooktitle{Woodstock '18: ACM Symposium on Neural Gaze Detection,
  June 03--05, 2018, Woodstock, NY}
\acmPrice{15.00}
\acmISBN{978-1-4503-XXXX-X/18/06}

\begin{document}

\title{Visual Encoding and Debiasing for CTR Prediction}

\author{
Si Chen$^\clubsuit$, Chen Lin$^\S$, Wanxian Guan$^\clubsuit$, Jiayi Wei$^\clubsuit$, Xingyuan Bu$^\clubsuit$, He Guo$^\clubsuit$, Hui Li$^\S$, Xubin Li$^\clubsuit$, Jian Xu$^\clubsuit$, Bo Zheng$^\clubsuit$ }   
\affiliation{\institution{$^\clubsuit$ Alibaba Group, Hangzhou }~~~~~~\institution{$^\S$ School of Informatics, Xiamen University}\country{China}}
\email{chenlin@xmu.edu.cn; lxb204722@alibaba-inc.com}

\renewcommand{\shortauthors}{Chen, et al.}

\begin{abstract}
Extracting expressive visual features is crucial for accurate Click-Through-Rate (CTR) prediction in visual search advertising systems. Current commercial systems use off-the-shelf visual encoders to facilitate fast online service. However, the extracted visual features are coarse-grained and/or biased. 
In this paper, we present a visual encoding framework for CTR prediction to overcome these problems. 
The framework is based on contrastive learning which pulls positive pairs closer and pushes negative pairs apart in the visual feature space. 
To obtain fine-grained visual features, we present contrastive learning supervised by click through data to fine-tune the visual encoder. To reduce sample selection bias, firstly we train the visual encoder offline by leveraging both unbiased self-supervision and click supervision signals. Secondly, we incorporate a debiasing network in the online CTR predictor to adjust the visual features by contrasting high impression items with selected items with lower impressions. 
We deploy the framework in the visual sponsor search system at Alibaba. Offline experiments on billion-scale datasets and online experiments demonstrate that the proposed framework can make accurate and unbiased predictions.   
\end{abstract}



\maketitle

\section{Introduction}
Visual search advertising systems, where products are displayed with images and images of products are accepted as queries, are a billion dollar business in E-commerce industry.  
Modern advertising systems use the cost-per-click marketing technique, which displays ads in search results whenever a user searches for a product, and gains revenue whenever the user clicks. 
The decision of ad placements is based on the product of predicted Click-Through-Rate (CTR) and the bid price. Therefore, to improve the performance of CTR prediction and consequently increase revenue, extracting expressive visual features is of vital importance. 

Current commercial visual search systems consist of two components~\cite{Liu2020Category}, the \VE with various CNNs
are trained \emph{off-the-shelf} to extract visual features, and the \CTR fuses visual features with non-visual features in different DNNs to make predictions. 
Challenges arise in training the \VE. 
On one hand, training the \VE with non-click-through signals leads to sub-optimal feature representations. 
For example, if the training task uses category labels, the feature representations can distinguish the category of clothes, but can not discover subtle style differences, which have a significant impact on user behaviors~\cite{Veit2015Learning}.   
On the other hand, if the \VE uses interaction signals such as clicks or purchases, it will face the problem of sample selection bias.  
For example, ads with low impressions (i.e., displayed less often in the system)  will receive fewer positive labels and therefore are under-represented in the learning process. 
 
This paper describes our solution in \Ali, which has one of China's largest E-commerce visual search platforms.
To facilitate real-time CTR prediction at scale, our solution also consists of two components, i.e., an off-the-shelf \VE and a \CTR.  
Our work is based on
contrastive learning, i.e., the learned feature representation of positive sample is pulled closer to the anchor image, while representation of a negative sample is pushed apart. 
\VE is firstly pre-trained with self-supervised contrastive loss, with random negative samples.
It is then fine-tuned with supervised contrastive loss, where the selection of positive and negative samples is dependent on user clicks.    
In this manner, we obtain finer-grained, more expressive visual features for CTR prediction, by leveraging user behavior information.  
In \CTR, we feed the extracted visual features through a \emph{debiasing network} before fusing with non-visual features. 
The debiasing network regularizes the CTR prediction loss with a contrastive loss, which encourages similar images from low impression items and high impression items to assemble.  
In this manner, we reduce the selection sample bias which has been introduced in the previous fine-tuning stage, while preserving the CTR prediction accuracy. 

In summary, our contributions are three-fold. 
(1) We study the problem of sample selection bias in visual features in advertising systems, which has not been explored in literature. Solutions to this problem shed light on the well-known ``accuracy-diversity'' dilemma in recommender systems.    
(2) We present a novel approach, which operates at Alibaba scale, to extract effective visual features for accurate and unbiased CTR prediction.  
(3) Offline experiments on ten-billion scale real production datasets demonstrate that pretraining-finetuning-debiasing has increased the accuracy of CTR prediction, especially for long-tail ads. 
Online A/B testing shows that, deploying the solution in Alibaba mobile app benefits the click-through rate and revenue per mille.

\section{Related Work}\label{sec:related}
Since CTR prediction is the central problem in online advertising industry, it has been extensively studied in academy and industry. 
Piorneer work~\cite{Chen2016Deep} extracts visual features of raw image and predicts CTR in one step. 
To speed up training online advertising system which encounters massive responses everyday, adopting off-the-shelf visual feature extraction modules has recently gained popularity ~\cite{He2016VBPR,Yang2019Learning,Kang2017Visually,Zhao2019What,He2016Sherlock,Liu2017DeepStyle,Liu2020Category,Zhao2017Photo2Trip,Yin2019Enhancing}. 
Most of them use CNNs as a visual encoder and pre-train the CNNs on image classification task. 
To learn visual compatibility across categories for fashion recommendation, the visual encoder in~\cite{Yin2019Enhancing} is pre-trained with weakly-labeled clothing collocation data.  
To learn category-specific inter-channel dependency, category-specific CNNs are adopted~\cite{Liu2020Category}. 
While images can be similar from multiple perspectives, training the visual encoder with image category labels is sub-optimal for CTR prediction. 
The click-through data is inheritantly biased, because ads must be exposured before being clicked.
However, to the best of our knowledge, SSB in visual feature extraction has not been explored. 
\section{Methodology}

\begin{figure*}[htbp]
\centering
\includegraphics[width=0.8\textwidth]{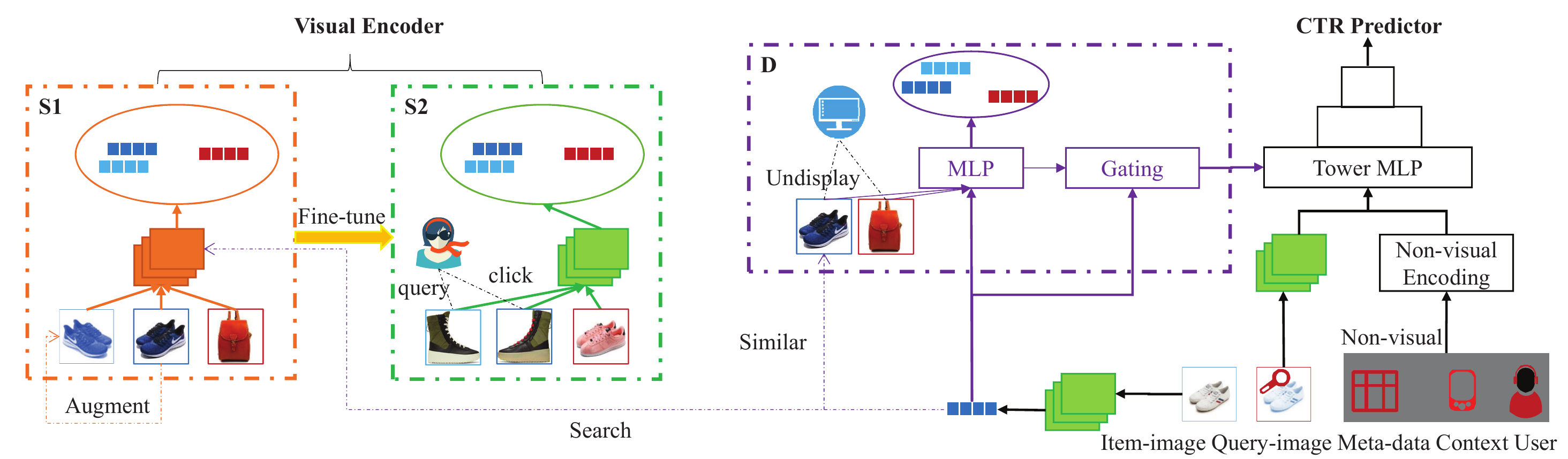}
\vspace*{-5pt}
\caption{The proposed architecture. Left: offline \VE consists of two stages. Right: online \CTR consists of a debiasing network. }\label{fig:framework}
\vspace{-10pt}
\end{figure*}
As shown in Figure~\ref{fig:framework}, the \VE  (Section~\ref{sec:ve}) extracts visual features for any image. It consists of two stages: \textbf{S1} and \textbf{S2}, both of which are based on contrastive learning. The \VE is trained offline separately, while the online serving system is the \CTR (Section~\ref{sec:ctr}). A debiasing network is plugged in \CTR to process visual features for ad items.  

\subsection{Visual Encoder}\label{sec:ve}
\textbf{S1: Pretraining \VE}. The standard self-supervised contrastive learning scheme is adopted. In a mini-batch of images $\mathcal{N}^{S1}$, for each anchor image $i\in\mathcal{N}^{S1}$, we augment it with a series of transformation, including random cropping, random color jitter, random greyscale, and random flipping. Thus, the positive sample $i'$ is obtained by $i'=t(i)$, where $t(\cdot)$ represents the transformation. The rest of the images within the mini-batch are considered as negative samples. Then, the anchor image, the positive sample, and the negative samples go through a visual encoder to obtain their visual features, by minimizing the contrastive loss:
\begin{equation}
\Loss_{S1}=-\sum_{i\in\mathcal{N}^{S1}} \log \frac{\exp\big(g(\vfi_i,\vfi_{t(i)}) \big)}{\sum_{j\in \mathcal{N}^{S1}\cup \{t(i)\}}\exp\big(g(\vfi_i,\vfi_{j}) \big) },
\end{equation}\label{equ:loss1}
where  $\vfi_i\in\Real^D$ is the output visual feature vector of image $i$, $D$ is the embedding size, $g(\vfi_i,\vfi_j)=cosine(\vfi_i,\vfi_j)$ is the cosine similarity between two visual feature vectors.  

\textbf{S2: Finetuning \VE}. After pretraining the visual encoder, we fine-tune its parameters. 
The difference between \textbf{S2} and \textbf{S1} lies in the construction of positive and negative samples. 

Clicks are one of the most invaluable sources to estimate visual relevance of an item given the query image. Thus we use the image of a clicked item as positive sample for an image query. However, it is well known that lack of clicks does not indicate irrelevance. To improve the quality of negative samples, we use the category information to build a negative sample pool. In E-commerce, each image is clearly labeled by its category (e.g., in the clothing section, an image could be labeled as ``dress'' or ``pants'', etc.).


For each query image $q$, we sample a clicked image $i$ as $q$'s positive image.
The category label of $i$ is denoted by $c_i$, $\mathcal{N}_{c_{i}}^{S2}$ is a collection of images of category label $c_i$ which can be seen as a negative sample pool.  
\begin{equation}
\Loss_{S2}=-\sum_{q\in\mathcal{Q}}  \log\frac{\exp\big(g(\vft_q,\vft_{i}) \big)}{\sum_{j\in \mathcal{N}_{c_{i}}^{S2} \cup \{i\}}\exp\big(g(\vft_{q},\vft_{j}) \big) },
\end{equation}\label{equ:loss2}
where $\vft_i\in\Real^D$ is the output visual feature of image $i$ in stage \textbf{S2}. It is of the same size as $\vfi_i$. $j\in\mathcal{N}_{c_{i}}^{S2}$ restricts negative samples belong to the same category as anchor, thus the negative samples are more informative and the contrastive task will be more difficult.
\subsection{CTR Predictor}\label{sec:ctr}
The \CTR aims to rank items in a pool of candidate ads to be displayed, by predicting the possibility of each item $p$ being clicked by user $u$ given query $q$ under context $x$. The inputs include the image of the item (to simplify notations, we also use $p$ to denote the item image), other item metadata such as item ID, shop ID, brand, category, price, and so on, user ID, user demographic features, preferred categories, and so on, context features such as device and position. Each query is an image, also denoted as $q$.
 
\textbf{Debiasing Network}. It is possible that \textbf{S2} introduces sample selection bias to the visual features. For example, longtail items with small impressions (i.e., number of times the ad has been displayed in total) are less likely to be clicked, and consequently make little contributions to S2. To eliminate such bias, in the \CTR, each item image goes through a debiasing network, which is also based on contrastive learning. 
Our intuition is to pull image-pairs that are visually similar but are significantly different in the number of impressions closer. In order to mine such sample pairs, we use unbiased S1 representation to depict the similarity of images and construct debiasing samples.

To construct positive sample for each anchor item image $p$,
we go through two steps. Firstly we retrieve a set $\tilde{\mathcal{P}}=\{p'\}$ of K
most similar images of non-displayed items with the same category label. We use the visual features extracted by stage S1 to compute the similarity, i.e., $sim(p,p')=cosine(\vfi_p,\vfi_{p'})$, so that the similarity will not be biased against longtail items. Secondly, the positive sample is selected based on the similarity, i.e., $Pr(p')=sim(p,p')/\sum_{p'\in\tilde{\mathcal{P}}} sim(p,p')$, where $Pr(p')$ is the probability of $p'$ being selected as a positive sample. 
The negative sample of each anchor is randomly selected. 

Then, the debiasing network \textbf{D} feeds a Multilayer Perceptron (MLP) with the visual features obtained by \textbf{S2}, i.e., $\vft_p$. The image $p$ is then contrasted positively with $p'$ and negatively with other images in the mini-batch $\mathcal{N}^{CTR}$. 
\begin{equation}
\Loss_{D}=-\sum_{p\in\mathcal{N}^{CTR}}\log\frac{\exp\big(g(\vfd_p,\vfd_{p'}) \big)}{\sum_{o\in\mathcal{N}^{CTR} \cup \{p'\}}\exp\big(g(\vfd_p,\vfd_{o}) \big) },
\end{equation}\label{equ:lossd}
where $\vfd_p\in\Real^D$ is the output visual feature of image $p$ in by the MLP, i.e., $\vfd_p=MLP(\vft_p)$. Minimizing $\Loss_D$ pushes item images with high impressions to be closer to similar item images with low impressions, and thus mitigates the bias of $\vft_p$.  

Next, $\vft_p$ and $\vfd_p$ go through a gating layer to generate effective and unbiased visual features for item $p$. 
$\alpha=\sigma\bigg(\mathbf{W}^T \big[ \vft_p,\vfd_p \big] \bigg),$
where $\sigma(\cdot)$ is the sigmoid function, $\mathbf{W}$ is a learnable weight matrix, $\big[\cdots\big]$ is the concatenation of several vectors/scalers
Finally, the visual feature of item $p$ is obtained: 
$\vf_p=\alpha \vft_p + (1-\alpha) \vfd_p.$

Since in this paper we focus on visual encoding, the rest of the \CTR can be very flexible as the pretraining-finetuning-debiasing network can plug into various frameworks. In the experiments, the visual feature of the query image $q$ is generated by the fine-tuned \VE, i.e., $\vf_q=\vft_q$.  The \CTR takes input of non-visual features, transforms them into embedding vectors through lookup tables, and feeds the concatenation of all embedding vectors to a tower MLP to make the prediction. Overall, the \CTR is optimized by minimizing the loss function:
\begin{equation}
\Loss_{CTR}= \Loss_{pred} +\Loss_{D}, 
\end{equation}\label{equ:lossall}      
where $\Loss_{pred}=-\sum_{y\in\mathcal{N}^{CTR} } \big[y\log(\hat{y}) + (1-y)\log(1-\hat{y})  \big] $ is the cross-entropy loss to evaluate CTR prediction accuracy, $y\in\{0,1\}$ is the actual click, and $\hat{y}$ is the predicted click probability. By incorporating $\Loss_D$, the debiasing network is trained jointly with \CTR to achieve accurate and unbiased predictions.

\section{Experiments}
In this section we analyze our experimental results in offline and online evaluations. The backbone of the visual encoder in S1 and S2 is ResNet50. We set the dimension size of visual features as $D=512$. In the debiasing network, we select $K=15$ similar images, the MLP has three hidden layers with $128, 16,128$ units, and the activation functions are $ReLU, tanh, ReLU$. The output layer has $512$ units to output the visual feature vector. The tower MLP in \CTR has three hidden layers with $512,256,128$ units, and the activation functions are $ReLU$, the output layer applies the sigmoid function to bound the prediction to $(0,1)$.  We use the Adagrad optimizer with learning rate $0.05$.



\subsection{Offline Visual Search Evaluation}
\textbf{Dataset.} To evaluate whether the extracted visual features are effective in identifying products, we perform a visual search task on an internal dataset. 
The dataset contains tens of thousands of 
item images sampled from multiple categories in our production system (e.g., clothing section, digital device section, furniture section, and so on.). The relevant image-pairs are manually annotated. The relevance judgement is binary (i.e., relevant or irrelevant), and it is based on a set of factors including style and design. 

\noindent\textbf{Baselines.} We compare the following visual encoders, including deep neural network classifiers and basic contrastive
learning methods. 
(1) ResNet-C: a ResNet50 is trained on the item images to predict the correct category labels. 
(2) S1: ResNet50 trained with self-supervised contrastive loss as in stage S1; 
(3) S2: the ResNet50 trained with click-through supervisions as described in stage S2; 
(4) S1+S2: first pretrain the ResNet50 as in stage S1 and then finetune it as in stage S2. 

\noindent\textbf{Evaluation Metric.}  After training each visual encoder $M$, visual feature vectors are extracted, we rank the images based on cosine similarity of visual feature vectors to the query image $q$. The result is denoted as $\mathcal{M}_q$. We adopt three evaluation metrics. (1) The primary metric is HitRatio, i.e., $HR=\sum_q |\mathcal{Q}_q\bigcap \mathcal{M}^{n_q}_q|/\sum_q |\mathcal{Q}^{n_q}_q|$, where $n_q$ is the number of relevant images in the groundtruth $|\mathcal{Q}_q|=n_q$. Higher $HR$ suggests higher search accuracy. (2) To reveal the diversity of results, we compute the ratio of images with low impressions in the returned images, i.e., $LR@K=\sum_q |\mathcal{L}\bigcap \mathcal{M}^K_q|/\sum_q |\mathcal{M}^K_q|$, where $\mathcal{L}$ is the set of images who receive less than five impressions during the last $30$ days, and $\mathcal{M}^K_q$ is the top-K results. Higher $LR@K$ suggests that the visual encoder is more fair to items with low impressions. (3) We also compute a supplementary metric, the ratio of images with the same categories in the results, i.e., $CR@K=\sum_q |\mathcal{C}_q\bigcap \mathcal{M}^K_q|/\sum_q |\mathcal{M}^K_q|$, where $\mathcal{C}_q$ is the set of images which are under the same category label of query image $q$. $CR$ provides information about the granularity of the visual features.
\begin{table}[]
\caption{Performance of visual search on manually annotated internal dataset: HitRatio $HR$, low-impression ratio $LR@K$, and same-category ratio $CR@K$.}
\resizebox{0.9\columnwidth}{!}{
\begin{tabular}{|c|c|cc|cc|}
\hline
\multirow{2}{*}{Method} &\multirow{2}{*}{HR}                     & \multicolumn{2}{c|}{LR}                      & \multicolumn{2}{c|}{CR}                      \\
\cline{3-6} 
      &                & \multicolumn{1}{l|}{LR@10}              & LR@100               & \multicolumn{1}{l|}{CR@10}                & CR@100               \\
       \hline
ResNet-C    & \multicolumn{1}{r|}{0.2626}  & \multicolumn{1}{r|}{0.5126} & \multicolumn{1}{r|}{0.5123} & \multicolumn{1}{r|}{0.8132} & \multicolumn{1}{r|}{0.7791} \\\hline
S1    & \multicolumn{1}{r|}{0.8504} & \multicolumn{1}{r|}{0.5001} & \multicolumn{1}{r|}{0.5059} & \multicolumn{1}{r|}{0.6357} & \multicolumn{1}{r|}{0.5524} \\\hline
S2    & \multicolumn{1}{r|}{0.8510} & \multicolumn{1}{r|}{0.5184} & \multicolumn{1}{r|}{0.5174} & \multicolumn{1}{r|}{0.7178} & \multicolumn{1}{r|}{0.6318} \\\hline
S1+S2    & \multicolumn{1}{r|}{\textbf{0.8825}} & \multicolumn{1}{r|}{\textbf{0.5259}} & \multicolumn{1}{r|}{\textbf{0.5207}} & \multicolumn{1}{r|}{0.7540$^{*}$} & \multicolumn{1}{r|}{0.6850$^*$} \\\hline
\end{tabular}}\label{tab:offlinevisual}
\end{table}
\vspace{-3pt}
\noindent\textbf{Analysis.} As shown in Table~\ref{tab:offlinevisual}, the proposed off-the-shelf visual encoding framework (i.e., S1+S2) achieves both highest accuracy (i.e., HR) and highest coverage of low impression items (i.e., LR). It outperforms using only self-supervision and click signals (i.e., S1 and S2 alone) in terms of all metrics, because the pretraining-finetuning framework adopts click-through data to obtain finer-grained features, and the self-supervision mitigates bias in click-through data. Although the conventional ResNet Classifier produces the highest CR, its HR is the lowest, which suggests that using category labels as supervision is able to capture coarse-grained category specific features but fails to capture fine-grained details such as style and design.  

\subsection{Offline CTR Evaluation}
\textbf{Dataset.} The offline CTR evaluation is conducted on a billion-scale Taobao dataset, which is collected from our production system, the training data spans for a period of 15 days sampled from July, 2021, 
with $0.4$ billion different item images and $1$ billion samples, and the testing data is collected from the next day of the last training date. 

\noindent\textbf{Evaluation protocols.} The competitors are CTR predictors using different visual encoding modules, including (1) ResNet-C, (2) VGG trained with category labels~\cite{Simonyan2015Very}, (3) VIT trained with category labels~\cite{Dosovitskiy2021Image}. We also conduct ablation study with different combinations of S1, S2, and D (debiasing network). The evaluation metric is AUC. We report the average AUC results and the AUC results on items with the lowest impressions (bottom $10\%$) and the highest impression (top $10\%$).

\begin{figure}
    \centering
    \includegraphics[width=0.9\linewidth]{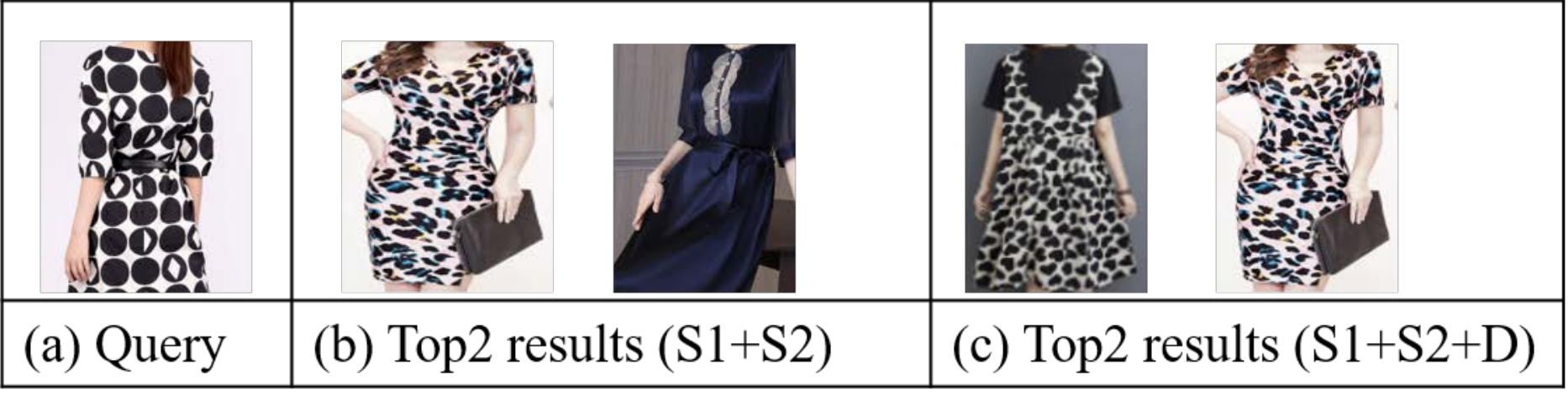}
    \vspace*{-5pt}
    \caption{A case study of debiased item ranking}
    \label{fig:case}
\end{figure}

\noindent\textbf{Analysis.} As shown in Table~\ref{tab:offlinectr}, compared with the best competitor ResNet, the proposed framework S1+S2+D increases AUC on testing set by $5\%$. Given the scale of our data, this is a significant improvement. Comparing among the different combinations of pretraining, finetuning and debiasing, we can see that neither S1 nor S2 alone can achieve optimal results. Furthermore, although S1+S2 can already produce good predictions, with the debiasing network, S1+S2+D is able to further improve predictions on low impression items over S1+S2, while preserving the overall accuracy for all items. We demonstrate the necessity of adopting the debiasing network by a case study in Figure~\ref{fig:case}, where the second result of S1+S2 is a more popular but not similar item, while S1+S2+D reduces bias against low impression items.  
\begin{table}[]
\caption{AUC of CTR prediction on Taobao dataset}
\vspace{-3pt}
\resizebox{0.9\columnwidth}{!}{
\begin{tabular}{|c|c|rrr|}
\hline
\multirow{2}{*}{Method}  & \multicolumn{1}{c|}{Visual Encoding} & \multicolumn{3}{c|}{Test}                                                                                  \\ \cline{2-5} 
      & Impression       & \multicolumn{1}{c|}{Bottom$10\%$}             & \multicolumn{1}{c|}{Top$10\%$}            & \multicolumn{1}{c|}{Overall} \\ \hline
\multirow{3}{*}{Competitors} & ResNet-C                      & \multicolumn{1}{r|}{0.7061}                & \multicolumn{1}{r|}{0.6959}                & 0.7042                       \\ \cline{2-5}
     & VGG                      & \multicolumn{1}{r|}{0.6667}                & \multicolumn{1}{r|}{0.6679}                & 0.6779                       \\ \cline{2-5}
     & VIT                     & \multicolumn{1}{r|}{0.7047}                & \multicolumn{1}{r|}{0.6971}                & 0.6981                       \\ \hline
\multirow{4}{*}{Ablation} & S1      & \multicolumn{1}{r|}{0.6874}         & \multicolumn{1}{r|}{0.6942}          & 0.7034                      \\\cline{2-5}
    & S2                       & \multicolumn{1}{r|}{0.7340}          & \multicolumn{1}{r|}{0.7240}          & 0.7293                       \\ \cline{2-5}
   & S1+S2                    & \multicolumn{1}{r|}{0.7673}          & \multicolumn{1}{r|}{0.7494}          & 0.7515                       \\ \cline{2-5}
 & S1+S2+D           & \multicolumn{1}{r|}{\textbf{0.7681}} & \multicolumn{1}{r|}{\textbf{0.7495}} & \textbf{0.7518}              \\ \hline
\end{tabular}\label{tab:offlinectr}}
\end{table}
\vspace{-3pt}
\vspace{-3pt}
\subsection{Online CTR Evaluation}
\begin{figure}[]
\centering
\includegraphics[width=0.9\columnwidth]{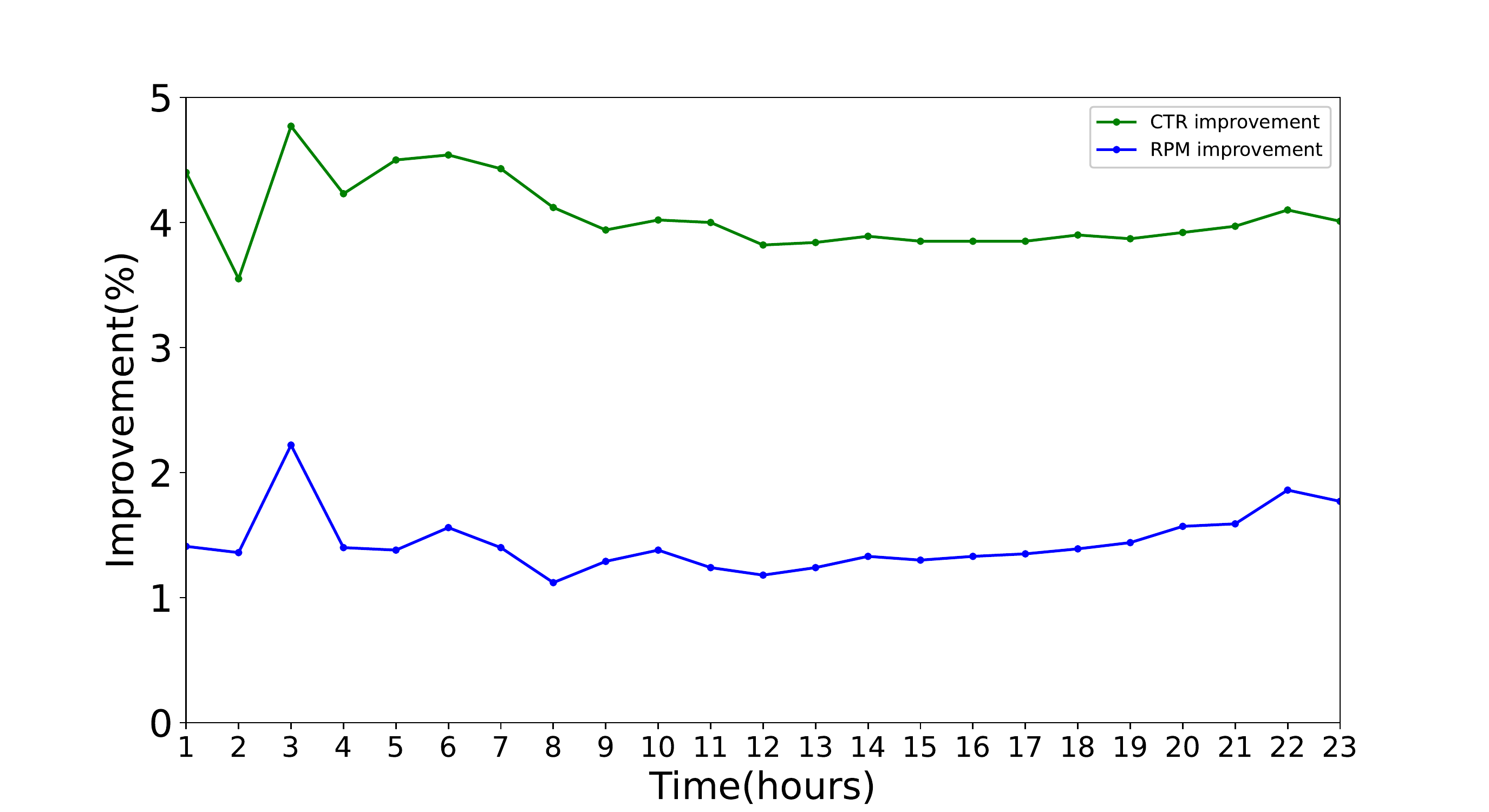}
\vspace*{-5pt}
\caption{CTR and RPM improvements during A/B testing}
\label{fig:online}
\vspace{-5pt}
\end{figure}
Finally we conduct an online A/B testing on the visual sponsor search system of Alibaba mobile application. The items of the control group in the A/B test period are provided by the previous version of online ranking system, which is based on S2 for visual encoding. The items offered to the experiment group are ranked based on the visual encoding S1+S2+D. We report the performance during 24 hours of the A/B test period. In Figure~\ref{fig:online}, $x$ represents the hours in a day, $y$ represents the CTR improvements and RPM (Revenue per Mille) improvements of the proposed framework with respect to the previous version up to this hour. We observe stable and significant increase of CTR ($4\%\sim 5\%$) and RPM ($1\%\sim 2\%$). 

\section{Conclusion}
This paper presents a pretraining-finetuning-debiasing framework to extract fine-grained and unbiased visual features for CTR prediction. The proposed system has been deployed online and powers the visual search advertising app at \Ali. We hope our experience helps commercial applications by more effective visual encoding.

\end{document}